\documentclass[useAMS,usenatbib]{mnras}
\usepackage{epsfig}
\usepackage{amsmath, amssymb,bm}
\usepackage[varg]{txfonts}
\usepackage{color}
\usepackage{aas_macros}

\title[Kozai--Lidov oscillations of circumplanetary discs]{Kozai--Lidov oscillations triggered by a tilt instability of detached circumplanetary discs}

\author[R. G. Martin et al.]{Rebecca G. Martin$^1$\thanks{E-mail:
    rebecca.martin@unlv.edu}, Zhaohuan Zhu$^1$, Philip J. Armitage$^{2,3}$,
   Chao-Chin Yang$^1$ \newauthor and Hans Baehr$^1$
\\$^1$Department of Physics and Astronomy, University of Nevada, Las
Vegas, 4505 South Maryland Parkway, Las Vegas, NV 89154, USA \\ 
$^2$Center for Computational Astrophysics,
Flatiron Institute, New York, NY 10010, USA\\
$^3$Department of Physics and
Astronomy, Stony Brook University, Stony Brook, NY 11794, USA}

\date{}

\pagerange{\pageref{firstpage}--\pageref{lastpage}} 
\pubyear{2021}

\begin{document}
\maketitle
\label{firstpage}
\begin{abstract} 
Circumplanetary discs can be linearly unstable to the growth of disc 
tilt in the tidal potential of the star-planet system. We use three-dimensional hydrodynamical simulations to characterize the disc conditions needed for instability, together with its long term evolution. Tilt growth occurs for disc aspect ratios, evaluated near the disc outer edge, of $H/r\gtrsim 0.05$, with a weak dependence on viscosity in the wave-like regime of warp propagation. Lower mass giant planets are more likely to have circumplanetary discs that satisfy the conditions for instability. We show that the tilt instability can excite the inclination to above the threshold where the circumplanetary disc becomes unstable to Kozai--Lidov (KL) oscillations. Dissipation in the Kozai--Lidov unstable regime caps further tilt growth, but the disc experiences large oscillations in both inclination and eccentricity. Planetary accretion occurs in episodic accretion events. We discuss implications of the joint tilt--KL instability for the detectability of circumplanetary discs, for the obliquity evolution of forming giant planets, and for the 
formation of satellite systems.
\end{abstract} 
  
 
\section{Introduction}   

A planet that forms in a protoplanetary disc is able to tidally open a gap in the protoplanetary disc when it reaches the mass of
Neptune
\citep{LP1986,DAngelo2002,Bate2003,Ayliffe2009,Ayliffe2009b}. Material
continues to flow from the circumstellar disc into the gap towards the
planet \citep{Artymowicz1996}. Since the planet has a size that is much smaller
than the Hill radius, a circumplanetary disc forms  \citep{Lubow1999,DAngelo2002}. The study of circumplanetary
discs is important for a number of reasons. First, most of the mass of
Jupiter may have been accreted from a circumplanetary disc. Thus, the
dynamics of a circumplanetary disc may have a major influence on
the properties of the forming planet. Secondly, regular satellites, those that have orbits that
are close to coplanar to the equatorial plane of the planet and low
orbital eccentricities, are thought to form in circumplanetary discs
\citep[e.g.][]{Lunine1982,Canup2002,
  Mosqueira2003,Batygin20}. Finally, circumplanetary discs may provide
observational signatures of forming planets \citep{Zhu2015b}.

Misaligned circumplanetary discs may change the obliquity of a forming
planet through both the accretion of misaligned material and the
torque acting to align the planet spin to the disc \citep{MZA2020}. In the solar system, Saturn, Uranus and Neptune have large obliquities and their regular satellites and ring systems are almost
aligned with the spin of each planet.  Processes that occur after satellite formation can also
lead to planet spin-orbit misalignment such as giant impacts
\citep{Safronov1966,Benz1989,Morbidelli2012}, spin-orbit resonances
\citep{Ward2004,Vokrouhlicky2015,Brasser2015,Rogoszinski2020} and
planet-circumstellar disc interactions \citep[][]{Millholland2019}.   A planet with a misaligned circumplanetary disc or a ring system 
would have deep transit. Some very low density
planets that have been observed in this way \citep{Masuda2014,JontofHutter2014} could be planets with
misaligned discs or rings \citep{Piro2020,Akinsanmi2020}.

In \cite{MZA2020} we found that the large disc aspect ratio and tidally truncated
size of circumplanetary discs means they are unstable to tilting through interaction with the tidal potential. 
\cite{Lubow92} first identified tilt
instabilities in tidally distorted discs. We assume that 
accretion onto the circumplanetary disc \citep[that has been examined by, for example by][]{Tanigawa12,Szulagyi2014,Schulik2020} can be ignored. This may occur, first, at the end of the lifetime of the circumstellar disc, when there is no further replenishment as the circumstellar disc has been accreted on to the star or photoevaporated \citep[e.g.][]{Clarke2001,Alexander2006,Owen2011}.  Secondly, this may occur if a wide gap is induced  by a massive planet in an inviscid disc \citep[e.g.][]{LP1986} or in a system with multiple planets \citep[e.g.][]{Zhu2011}.  Stochastic accretion of gas from the turbulent protoplanetary disc \citep{Gressel2013} may provide the small initial tilt that is needed for the disc to be unstable to tilting.

In this work, we study the interaction between the tilt instability and 
Kozai--Lidov \citep[KL,][]{Kozai1962,Lidov1962} oscillations of
inclination and eccentricity, which can occur in both point-mass \citep{Naoz2016} and fluid disc \citep{Martinetal2014b,Fu2015,Fu2015b,Franchini2019} systems. KL dynamics become important above a critical inclination. For a test particle this inclination is about $39^\circ$ but for a disc  it depends upon the disc aspect ratio
\citep{Lubow2017,Zanazzi2017}. Moderately tilted discs may then see their tilts increase due to the tilt instability, before entering a regime in which both the tilt and KL instabilities are active. In Section~\ref{sph} we discuss the
results of SPH simulations of detached circumplanetary discs. In Section~\ref{longterm} we discuss some effects that were neglected in our simulations and their effects on the evolution. In
Section~\ref{discussion} we discuss implications of our results on
giant planet formation, regular satellite formation and observations
of forming planets. We conclude in Section~\ref{concs}.

\section{Hydrodynamical simulations}
\label{sph}

We use the smoothed particle hydrodynamics (SPH) code {\sc Phantom}
\citep{PF2010,LP2010,Price2018} to model a detached circumplanetary
disc that is misaligned to the planet orbit. {\sc Phantom } has been
used extensively to model misaligned discs in binary systems
\citep[e.g.][]{Nixon2012,Nixonetal2013,Smallwood2019,Franchini2020,Bi2020}. There
are two sink particles \citep[e.g.][]{Bate1995}, one representing the
star with mass $M_{\rm s}$ and the other representing the planet with
mass $M_{\rm p}=10^{-3} \,M_{\rm s}$. The simulations do not 
depend on the orbital separation of the planet, $a$, because all lengths
are scaled to the Hill radius given by
\begin{equation}
r_{\rm H}=a\left(\frac{M_{\rm p}}{3M_{\rm s}}\right)^{1/3}.
\end{equation}
We choose the accretion radius of the star to be $1.4\,r_{\rm H}$, and that of
the planet is $0.03\,r_{\rm H}$ \citep{MZA2020}. Any particle that moves within the
accretion radius is accreted on to the sink. The mass and angular
momentum of the accreted particle are added to the sink so that
  mass, linear momentum and angular momentum are conserved.  The
planet is in a circular orbit with orbital period $P_{\rm orb}=2\pi/\sqrt{G(M_{\rm p}+M_{\rm s})/a^3}$.

The simulation parameters are summarised in Table~\ref{table}. The
surface density of the disc is initially distributed as 
$\Sigma \propto r^{-3/2}$ between the inner radius of $r_{\rm in}=0.03\,r_{\rm H}$ up to the outer radius of
$r_{\rm out}=0.4\,r_{\rm H}$.  Because the density evolves
quickly, the initial truncation radii of the disc
do not significantly affect the disc evolution. The initial outer truncation radius is approximately the size of a tidally
truncated circumplanetary disc \citep{Martin2011}, although the
truncation radius of a misaligned disc may be larger
\citep{Lubow2015,Miranda2015}. There is no addition of material to the
disc during the simulation.  We do not include disc self-gravity and so the mass of the disc does not affect the disc dynamics.  We discuss this assumption further in Section~\ref{longterm}. The initial
total disc mass is small compared to the planet mass, $M_{\rm
  d0}=10^{-6}\,M_{\rm s}$. This is about the maximum steady state disc
mass found in two-dimensional hydrodynamic simulations for various disc parameters with infall accretion rates up to $10^{-9}\,\rm M_\odot \, yr^{-1}$
\citep{Chen2021}. The tidal torque from the Sun excites spiral
density waves that result in shocks that transport angular momentum 
through the disc even in the absence of other sources of viscosity \citep{Zhu2016}.

The disc is initially inclined to the orbital plane of the planet with
inclination $i_0$.  We consider three different initial tilt angles, $i_0=10^\circ$, $i_0=30^\circ$ and $i_0=60^\circ$. Small tilt angles, up to about $15^\circ$ may be expected during the formation of a circumplanetary disc due to stochastic accretion flow into the planet Hill sphere \citep{Gressel2013}. While we do not expect the disc to form with larger initial tilt, we consider larger values that may be achieved because of the tilt instability \citep{MZA2020}. The long term effects and some caveats are discussed in Section~3 in more detail.

The disc is locally isothermal with fixed sound speed
$c_{\rm s}\propto r^{-q}$. Typically we take $q=0.75$ so that $\alpha$
and the smoothing length $\left<h\right>/H$ are constant over the disc
\citep{LP2007} although we consider one simulation with $q=0.5$
(constant disc aspect ratio).  We generally take the \cite{SS1973}
$\alpha$ parameter to be 0.01 although we consider one simulation with
a smaller $\alpha=0.005$.  The disc viscosity is implemented by adapting
the SPH artificial viscosity according to the procedure described in
\cite{LP2010} with $\alpha_{\rm AV}$ shown in column~5 of Table~1 and
$\beta_{\rm AV}=2$. The circumplanetary disc is initially resolved
with shell-averaged smoothing length per scale height shown in column~4 of
Table~1.

In order to analyse the simulations we divide the particles into 100
bins that are linear in spherical radius. The bins extend from
the innermost particle around the planet to the outermost particle
that is bound to the planet. Within each bin we calculate the mean values for the surface density, $\Sigma$, the disc inclination
with respect to the planet orbital plane, $i$, the disc nodal phase
angle (longitude of the ascending node), $\phi$, the disc
eccentricity, $e$, and the disc argument of periapsis, $\phi_e$. 

We
define the tilt growth timescale of the disc as
\begin{equation}
t_{\rm growth}=\frac{i}{di/dt}.
\label{tgrowth}
\end{equation}
We approximate the growth timescale by calculating a least squares
quartic polynomial fit to the inclination of the disc in time. In
Table~1 we show the growth timescale at $t=10\,P_{\rm orb}$ for
simulations in which the disc is not undergoing KL oscillations at
that point. Simulations with an initial inclination of $i_0=10^\circ$
are fit up to time $t=40\,P_{\rm orb}$, simulations with an initial
inclination of $i_o=30^\circ$ are fit up to a time of $t=30\,P_{\rm
  orb}$, while simulations with an initial inclination of
$i_0=60^\circ$ are fit up to a time of $t=12\,P_{\rm orb}$. 
After these time intervals the discs may undergo KL oscillations.

\begin{figure*} 
\begin{centering} 
\includegraphics[width=5.6cm]{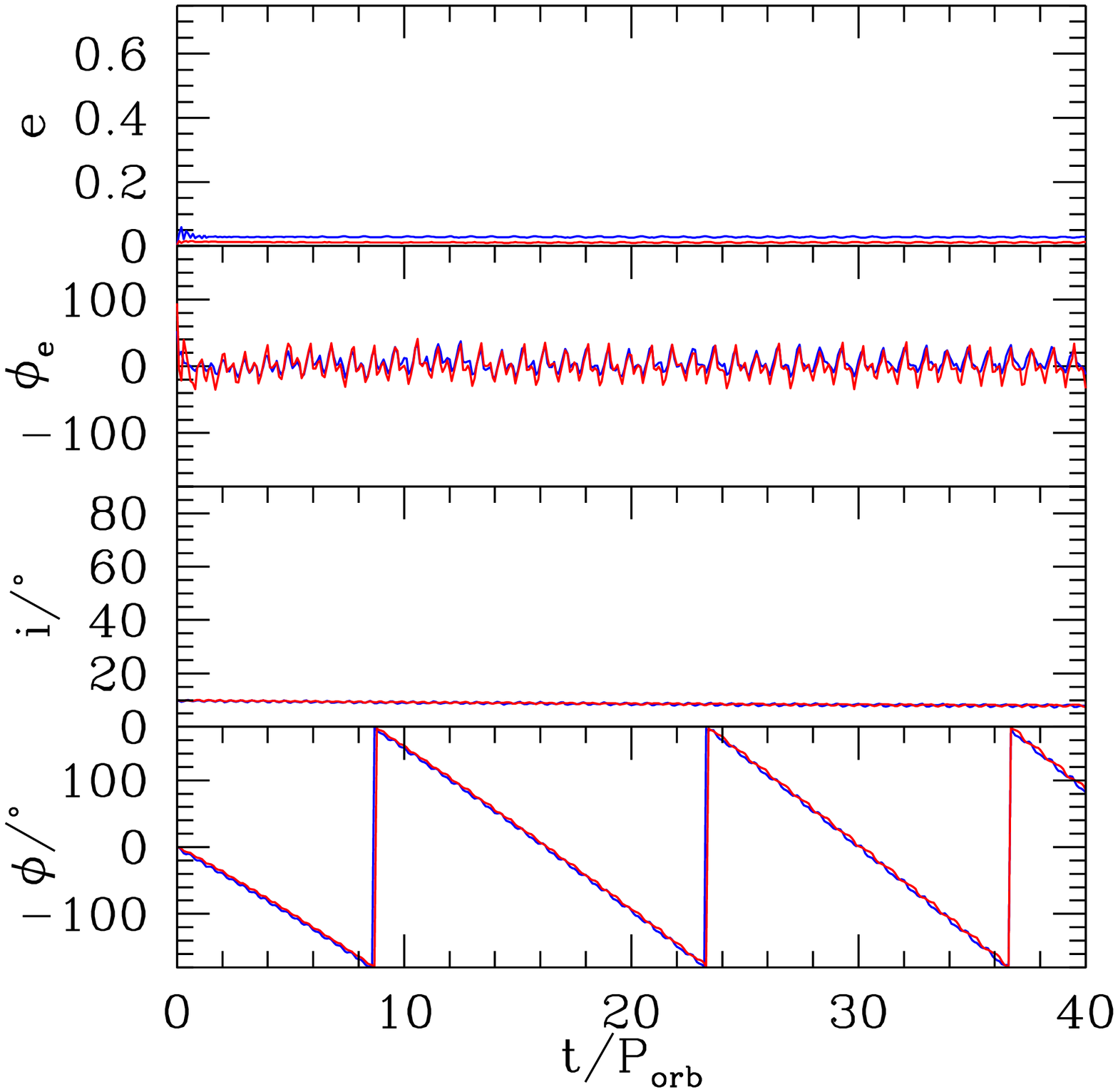} 
\includegraphics[width=5.6cm]{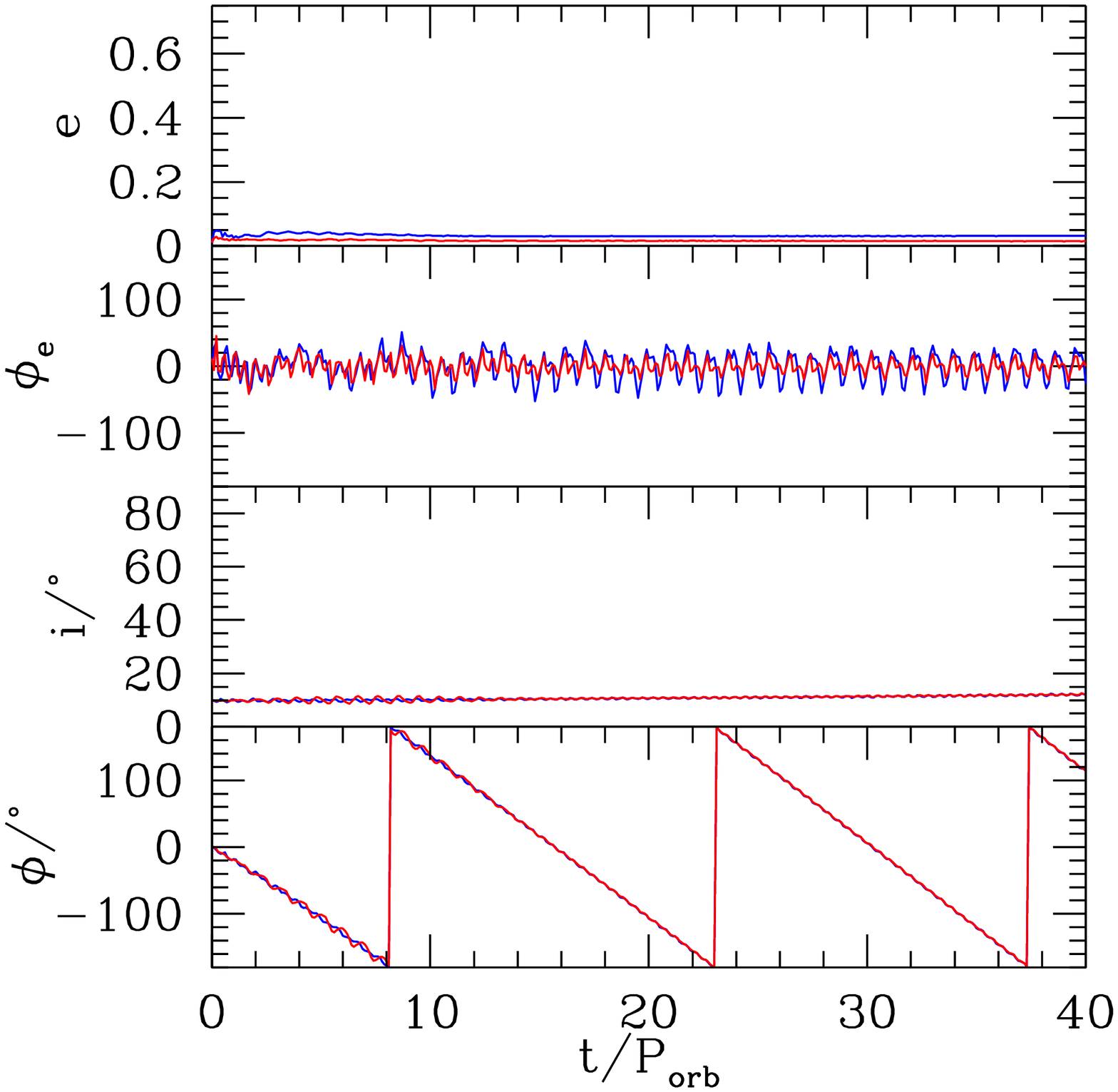} 
\includegraphics[width=5.6cm]{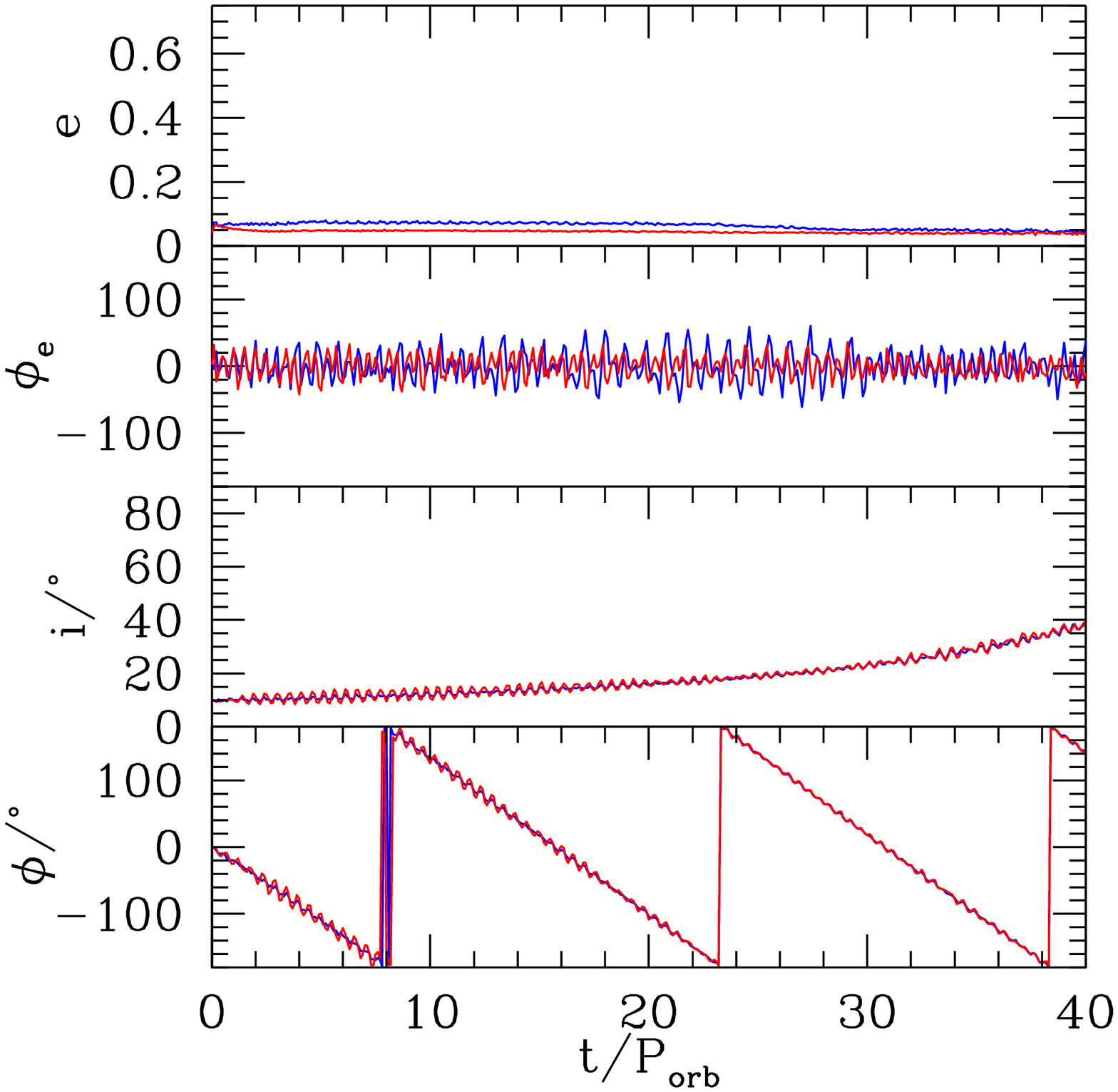} 
\end{centering} 
\caption{Hydrodynamical simulations of a circumplanetary disc with
  initial inclination of $i_0=10^\circ$ with disc aspect ratio
  $(H/r)_{\rm out}=0.025$ (left, run1), $(H/r)_{\rm out}=0.05$
  (middle, run2) and $(H/r)_{\rm out}=0.1$ (right, run3). The upper
  panel shows  the eccentricity of the disc, the
  second panel shows the argument of periapsis, the third panel shows
  the inclination of the disc and the lower panel shows the nodal
  phase angle. The red lines show the disc orbital properties at radius $r=0.2\,r_{\rm
    H}$ and the blue lines at $r=0.3\,r_{\rm H}$.}
\label{fig:inc10} 
\end{figure*}

\begin{table*}
\begin{center}
\begin{tabular}{lccccccccccc}
\hline
Simulation & Figure &$N$ & $\left<h\right>/H$ & $\alpha_{\rm AV}$ &$q$& $(H/r)_{\rm in}$ &$(H/r)_{\rm out}$ & $i_0/^\circ$ &  $t_{\rm growth}(10\,P_{\rm orb})/P_{\rm orb}$& KL  \\
\hline
\hline
run1 & \ref{fig:inc10}   & 500,000 & 0.60 & 0.17 & 0.75 & 0.05 & 0.025& 10  & -147 & No \\
run2 &  \ref{fig:inc10}  & 500,000 & 0.37 & 0.27 &  0.75 &0.1 &0.05 & 10  & 204 & No \\
run3 &  \ref{fig:inc10} & 1,000,000 & 0.17 & 0.58  &0.75 & 0.2    &    0.1  & 10  & 39  & No \\
\hline
run4   & \ref{fig:q0p5} & 500,000 & 0.25 & 0.39 & 0.5  & 0.1 & 0.1 & 10 &  51 & No \\
\hline
run5 & \ref{fig:lowinc} & 500,000 & 0.59 & 0.17 &0.75 &0.05 & 0.025 & 30  &  -232 & No\\
run6 & \ref{fig:lowinc} & 500,000 & 0.37 & 0.27  &   0.75  & 0.1 & 0.05 & 30  & 257  & No \\
run7 & \ref{fig:lowinc} & 1,000,000 & 0.19& 0.53 &0.75 & 0.2 & 0.1 & 30  & 43 & No$\rightarrow$ Yes \\
\hline
run8   & \ref{fig:highinc} & 500,000 &0.58& 0.17 & 0.75 &0.05 & 0.025 & 60 &  - & Yes \\
run9  & \ref{fig:highinc} & 500,000 &0.37 & 0.27  &0.75 &0.1 & 0.05 & 60 &  - & Yes\\
run10  & \ref{fig:highinc} & 1,000,000 & 0.18& 0.54  & 0.75 &0.2 & 0.1  & 60 & 72  & No $\rightarrow$ Yes \\
\hline
run11 & \ref{fig:alpha} & 1,000,000 & 0.18 & 0.27 & 0.75 & 0.2 & 0.1 & 60 & 80 & No $\rightarrow$ Yes \\
\hline
\end{tabular}
\end{center}
\caption{Parameters of the SPH simulations. The first column is the
  name of the simulation. The second column shows which Figure the
  simulation appears in. The third column is the initial number of
  particles in the simulation. The fourth column is the initial mass
  averaged smoothing length divided by the disc scale height. The
  fifth column is the artificial viscosity parameter. The sixth column
  is the sound speed index, $c_{\rm s}\propto r^{-q}$. The seventh
  column is the disc aspect ratio at the disc inner edge. The eighth
  column is the disc aspect ratio at the initial disc outer
  radius. The ninth column is the initial disc inclination. The tenth
  column shows the disc tilt growth timescale at time $t=10\, P_{\rm
    orb}$. The eleventh column describes whether the disc is unstable
  to KL oscillations. }
\label{table}
\end{table*}

\begin{figure*} 
\centering
\includegraphics[width=16.8cm]{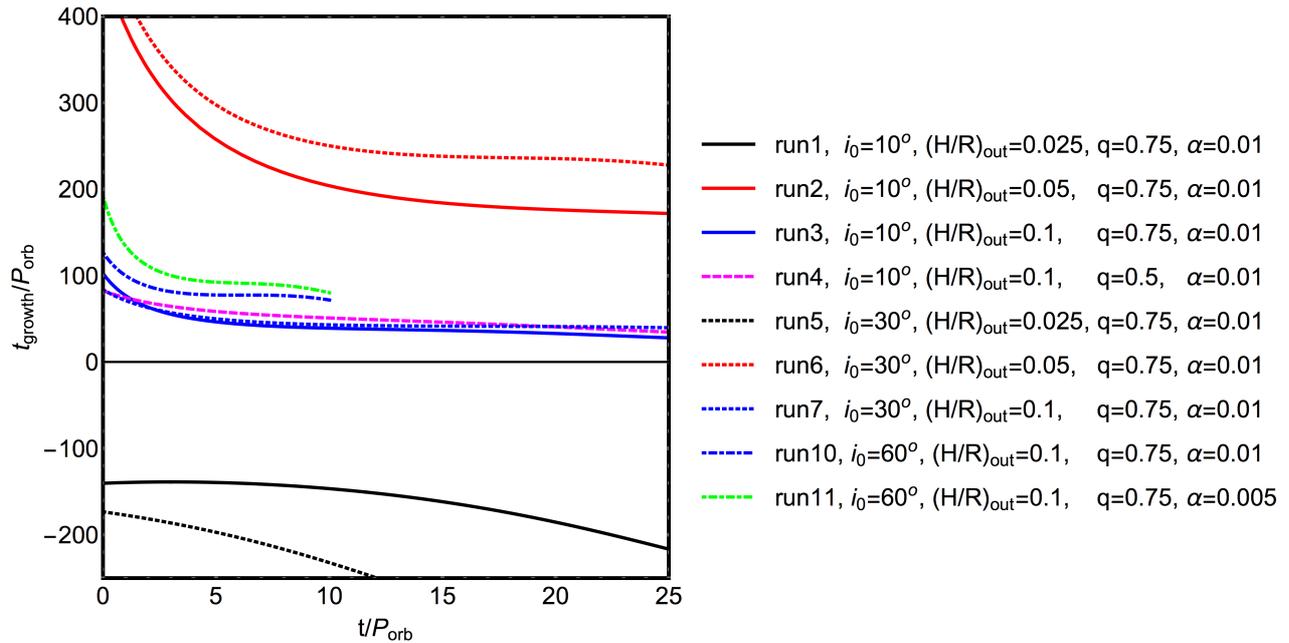} 
\caption{The approximate tilt growth timescale as a function of
  time for circumplanetary discs with initial misalignment
  $i_0=10^\circ$ (solid and dashed lines), $i_0=30^\circ$ (dotted
  lines) and $i_0=60^\circ$ (dot-dashed lines). All lines have
  $q=0.75$ except the magenta dashed line which has $q=0.5$ and
  $H/r=0.1$. All lines have $\alpha=0.01$ except the green line that
  has $\alpha=0.005$. The disc aspect ratio is $(H/r)_{\rm out}=0.025$
  (black lines), $(H/r)_{\rm out}=0.05$ (red lines) and $(H/r)_{\rm
    out}=0.1$ (blue, magenta and green lines). }
\label{doublingtimeplot} 
\end{figure*}

\begin{figure*} 
\centering
\includegraphics[width=16.8cm]{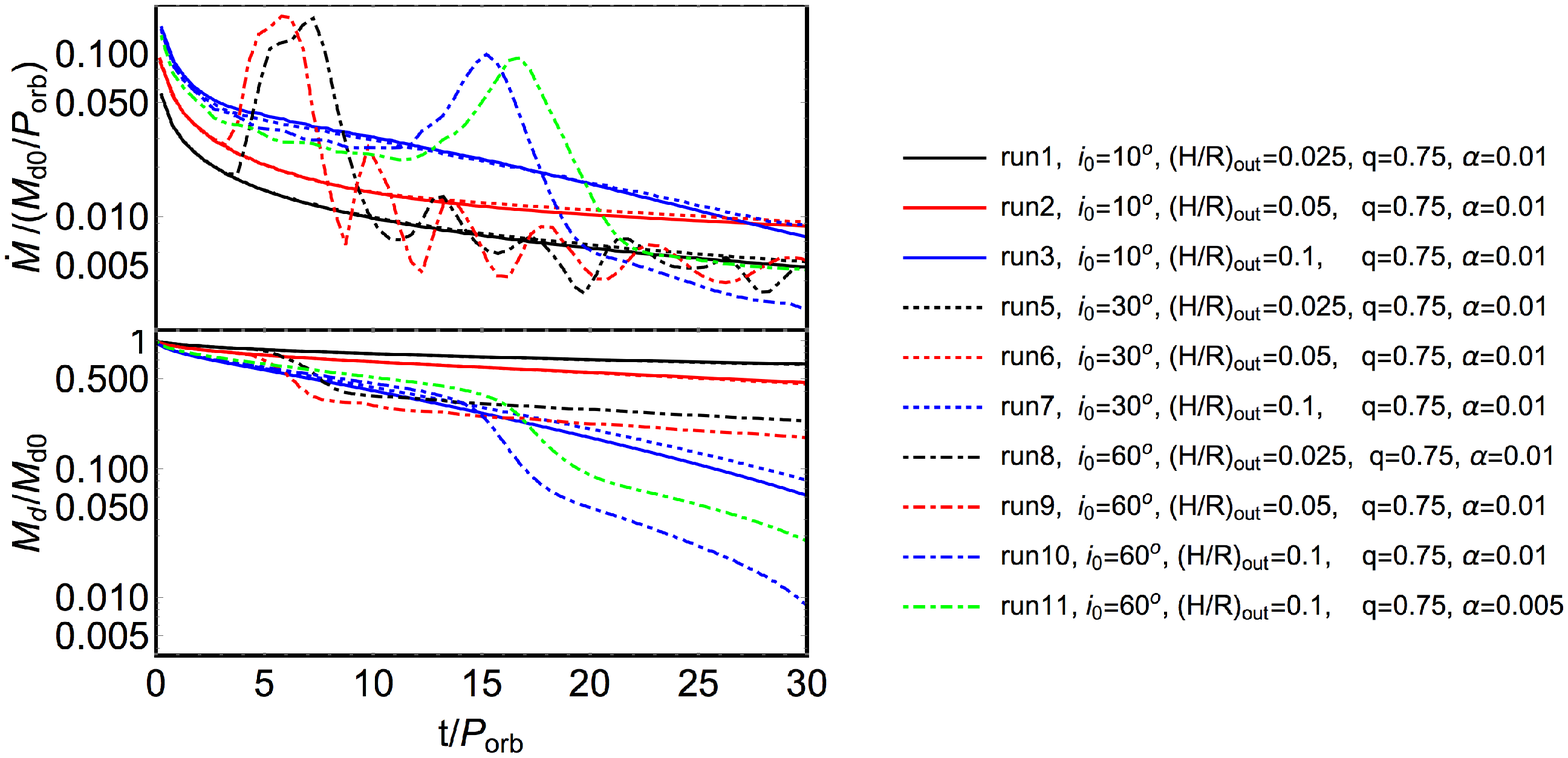} 
\caption{ Time averaged accretion rate on to the planet (upper panel)
  and the mass of the disc (lower panel). The
  black lines show simulations with $(H/r)_{\rm out}=0.025$, the red
  lines show simulations with $(H/r)_{\rm out}=0.05$ and the blue
  lines show simulations with $(H/r)_{\rm out}=0.1$.  All the
  simulations have $\alpha=0.01$ except the green dot-dashed line that
  has $\alpha=0.005$ and $(H/r)_{\rm out}=0.1$. The solid lines have
  initial inclination $i_0=10^\circ$, the dotted lines have
  $i_0=30^\circ$ and the dot-dashed lines have $i_0=60^\circ$. }
\label{acc} 
\end{figure*}  

\subsection{Mildly inclined circumplanetary discs}

Two components of the tidal potential determine the  dynamics of a mildly misaligned disc. Retrograde nodal precession is caused by the $m=0$ component. In the wave-like regime, where $\alpha < H/r$, the disc holds
itself together through wave-like communication \citep{Papaloizou83} and this leads to solid body nodal precession  \citep{PT1995,Larwoodetal1996,Terquem1998}. 
An ``oscillating'' torque with a period of half the orbital
period is produced by the $m=2$ component of the torque. This does not change the mean precession rate
\citep{Katz1982}. Dissipation within the disc means that  the $m=0$ component leads the disc to move towards coplanar alignment where the disc is aligned to the orbital plane of the planet. The $m=2$ term on the other hand leads to the tilt increasing \citep{Lubow92,Lubow2000,Bateetal2000}. For circumstellar
discs, the outcome is normally coplanar alignment. However, because of the tidally truncated size of
circumplanetary discs  and their typically larger disc aspect ratio,
their tilt tends to increase \citep{MZA2020}.

Fig.~\ref{fig:inc10} shows three SPH simulations that begin with an
initial disc inclination of $i_0=10^\circ$ for different disc aspect
ratios. The sound speed power law in each case has $q=0.75$ meaning
that $H/r\propto r^{-1/4}$. The value of $H/r$ at the initial disc
outer edge, $(H/r)_{\rm out}=H/r(r=r_{\rm out})$, is about half the
value at the inner edge, $(H/r)_{\rm in}=H/r(r=r_{\rm in})$.  Each
panel shows the disc eccentricity, the disc argument of periapsis, the
disc inclination and the disc nodal phase angle from top to
bottom. Within each panel the red lines show the evolution in the
middle of the disc at a radius $r=0.2\,r_{\rm H}$ while the blue lines
show the evolution farther out at $r=0.3\,r_{\rm H}$.

The left panel of Fig.~\ref{fig:inc10} shows the lowest disc aspect
ratio considered, $(H/r)_{\rm out}=0.025$ (run1). The disc in this
simulation nodally precesses at a roughly constant rate on a timescale
of about $18\,P_{\rm orb}$. The tilt shows a slow decay. Since the red
and the blue lines are very similar, the disc is precessing
globally. The middle panel shows a higher disc aspect ratio of
$(H/r)_{\rm out}=0.05$ (run2). The tilt grows in this simulation.
The right hand panel shows the SPH simulation presented in
\cite{MZA2020} that has $(H/r)_{\rm out}=0.1$ (run3).  The disc tilt
grows on a shorter timescale for larger disc aspect ratio.  There is
no eccentricity growth in any of these low initial tilt simulations.

Fig.~\ref{doublingtimeplot} shows the tilt growth timescale
(equation~\ref{tgrowth}) as a function of time. Positive values
represent discs with increasing tilts while negative values represent
those with decreasing tilts. The three solid lines show the mildly
misaligned simulations with $i_0=10^\circ$ and $q=0.75$ for the three
different disc aspect ratios $(H/r)_{\rm out}=0.025$ (run1, black),
$(H/r)_{\rm out}=0.05$ (run2, red) and $(H/r)_{\rm out}=0.1$ (run3,
blue).  For the simulations with increasing tilts (positive $t_{\rm
  growth}$), the growth timescale decreases over time but reaches a
quasi-steady value.

Fig.~\ref{acc} shows the time averaged accretion rate on to the planet
and the disc mass. The solid lines show the
simulations that are initially misaligned by $i_0=10^\circ$ with
$q=0.75$. The lower the disc aspect ratio the slower the accretion on
to the planet since the viscosity is lower. The accretion rate on to the planet
decreases in time because the mass in the circumplanetary disc decreases.

\subsubsection{Effect of the temperature structure}

\begin{figure} 
\centering
\includegraphics[width=5.6cm]{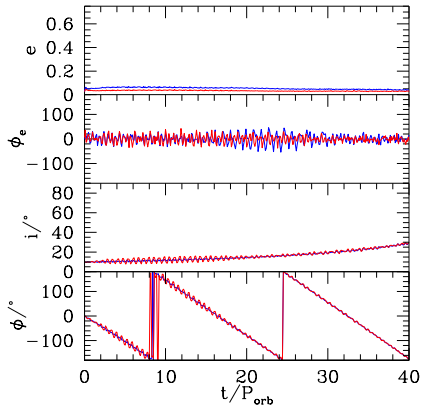} 
\caption{Same as the right panel of Fig.~\ref{fig:inc10} (run3)
  except $q=0.5$ (run4). Simulations run3 and run4 have the same
  disc aspect ratio $(H/r)_{\rm out}=0.1$. }
\label{fig:q0p5} 
\end{figure}

Fig.~\ref{fig:q0p5} shows the same simulation as in
Fig.~\ref{fig:inc10} but with a disc aspect ratio that is constant
with radius ($q=0.5$) and $H/r=0.1$ (run4). The disc evolution is
similar to the right hand panel of Fig.~\ref{fig:inc10} which has
$(H/r)_{\rm out}=0.1$ (run3). The dashed magenta line in
Fig.~\ref{doublingtimeplot} shows the tilt growth timescale for
the simulation with $i_0=10^\circ$ and $q=0.5$ is most similar to the
simulation with the same initial inclination and $(H/r)_{\rm out}$ but
$q=0.75$ (solid blue line, run3). Thus, the disc aspect ratio at the
outer disc edge is what determines the disc evolution (not the disc
aspect ratio in the inner parts of the disc) and the results are
relatively insensitive to $q$. For more realistic temperature
structures (than our standard choice) we expect the aspect ratio at
the outer disc edge to determine the tilt growth timescale.

\subsection{Moderately inclined circumplanetary discs}

\begin{figure*} 
\begin{centering} 
\includegraphics[width=5.6cm]{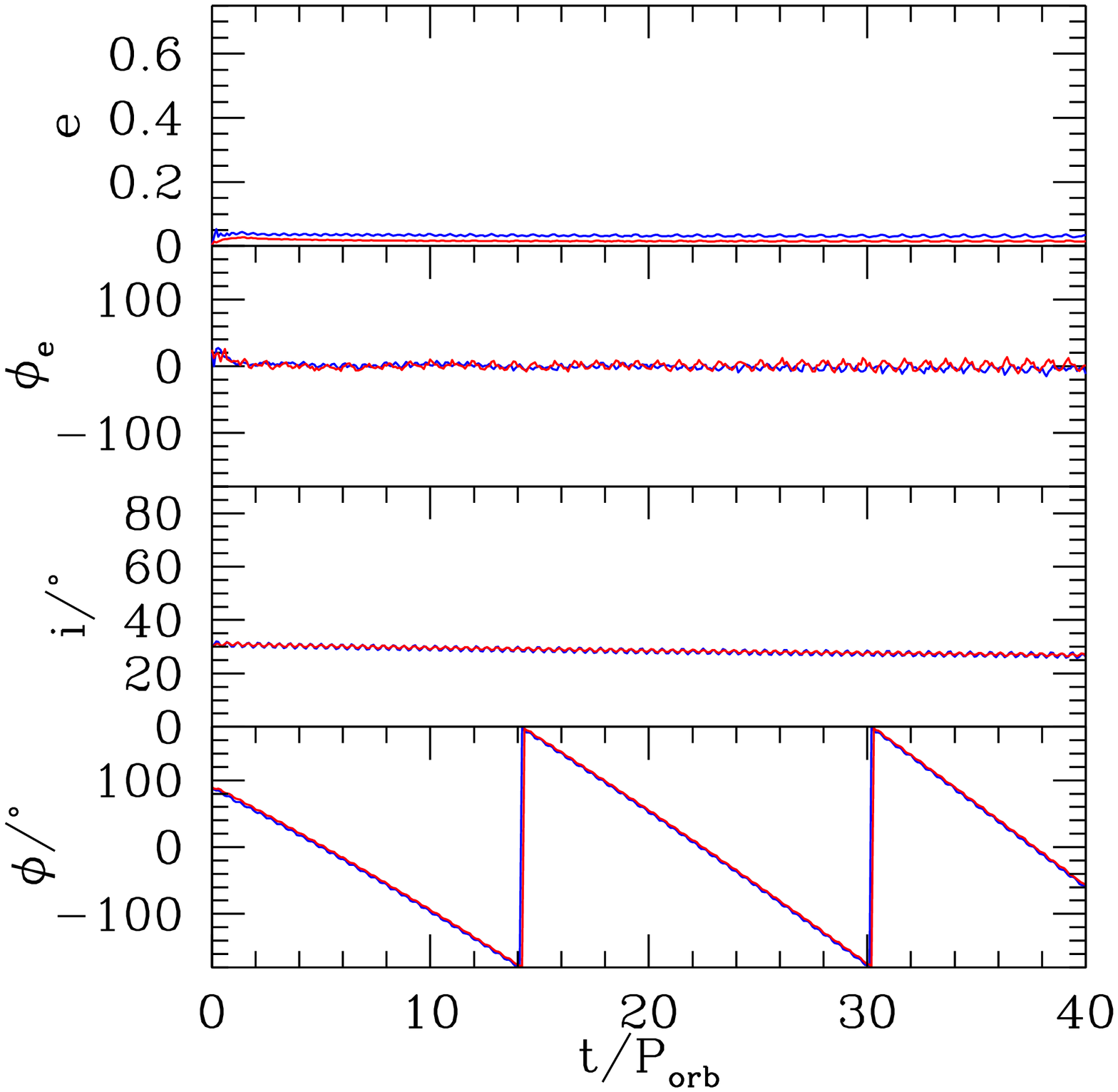} 
\includegraphics[width=5.6cm]{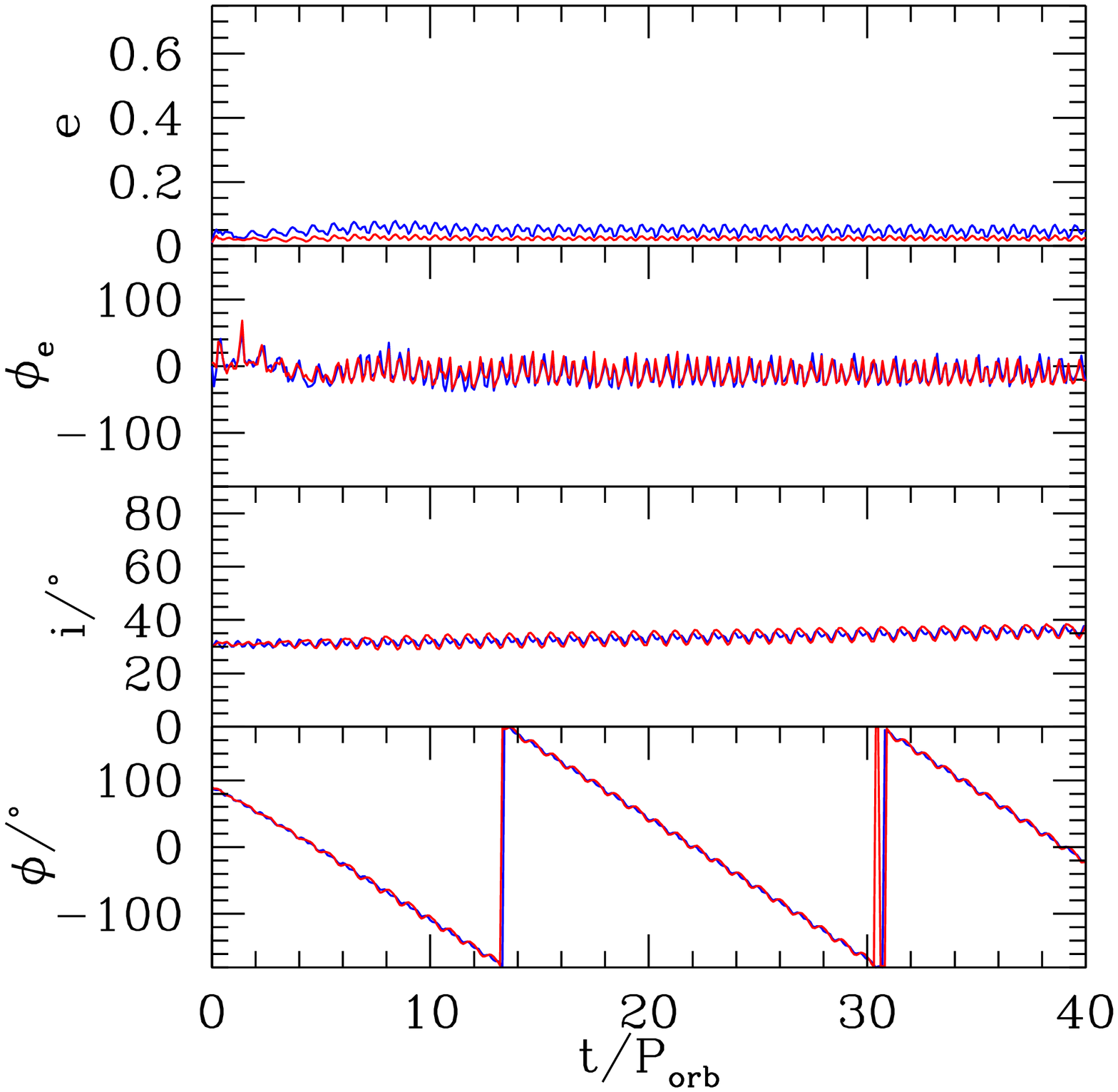} 
\includegraphics[width=5.6cm]{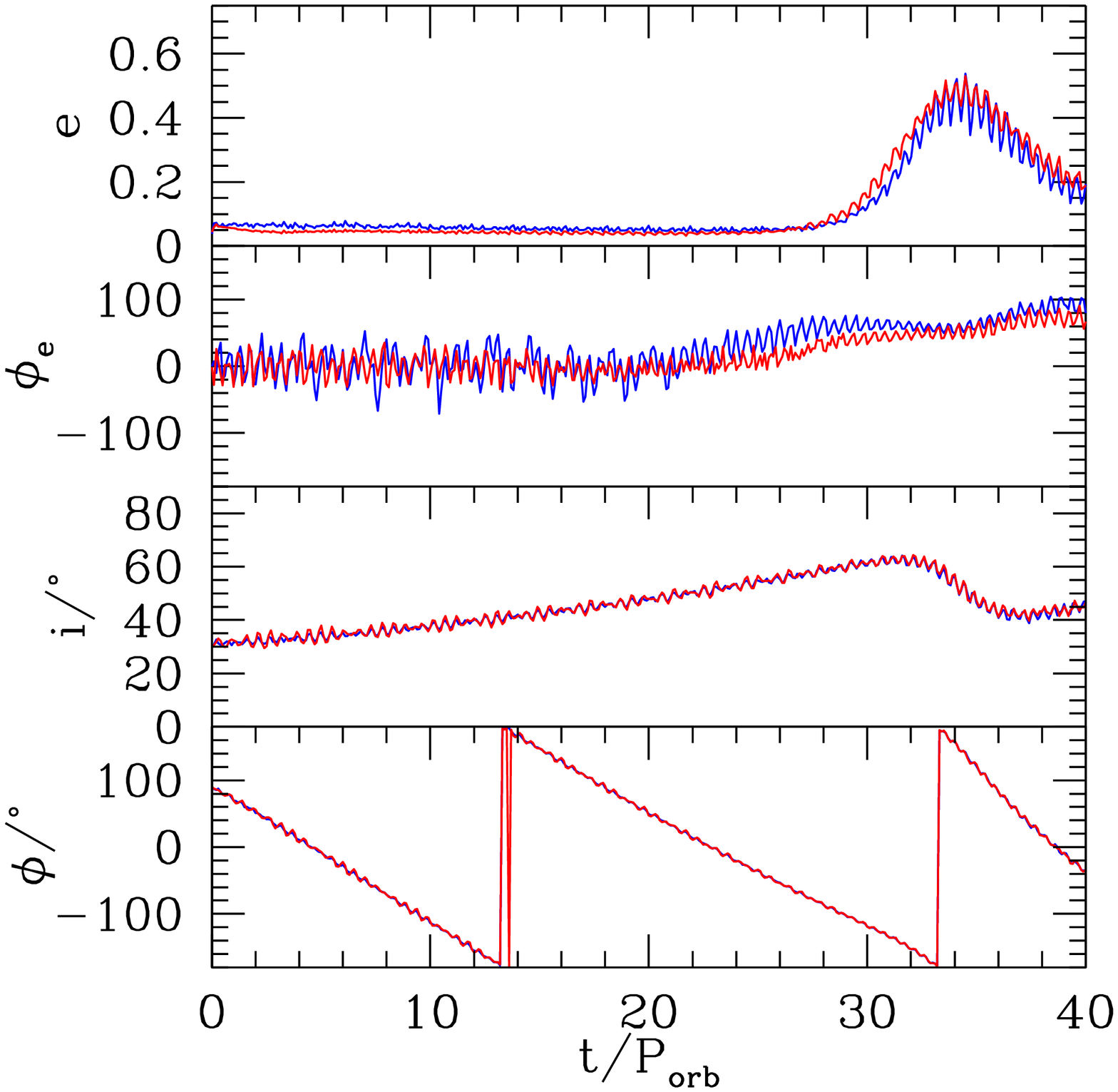} 
\includegraphics[width=5.6cm]{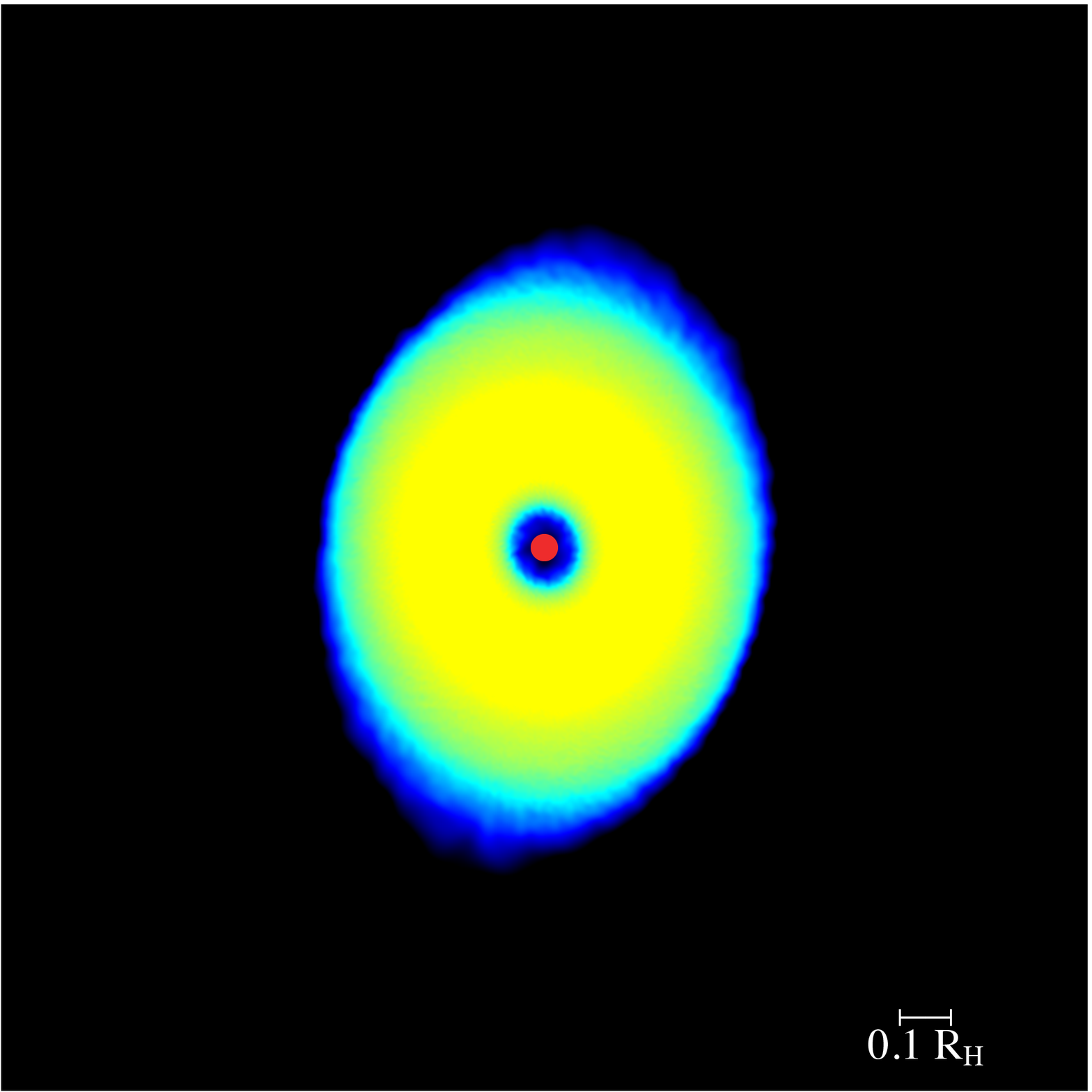} 
\includegraphics[width=5.6cm]{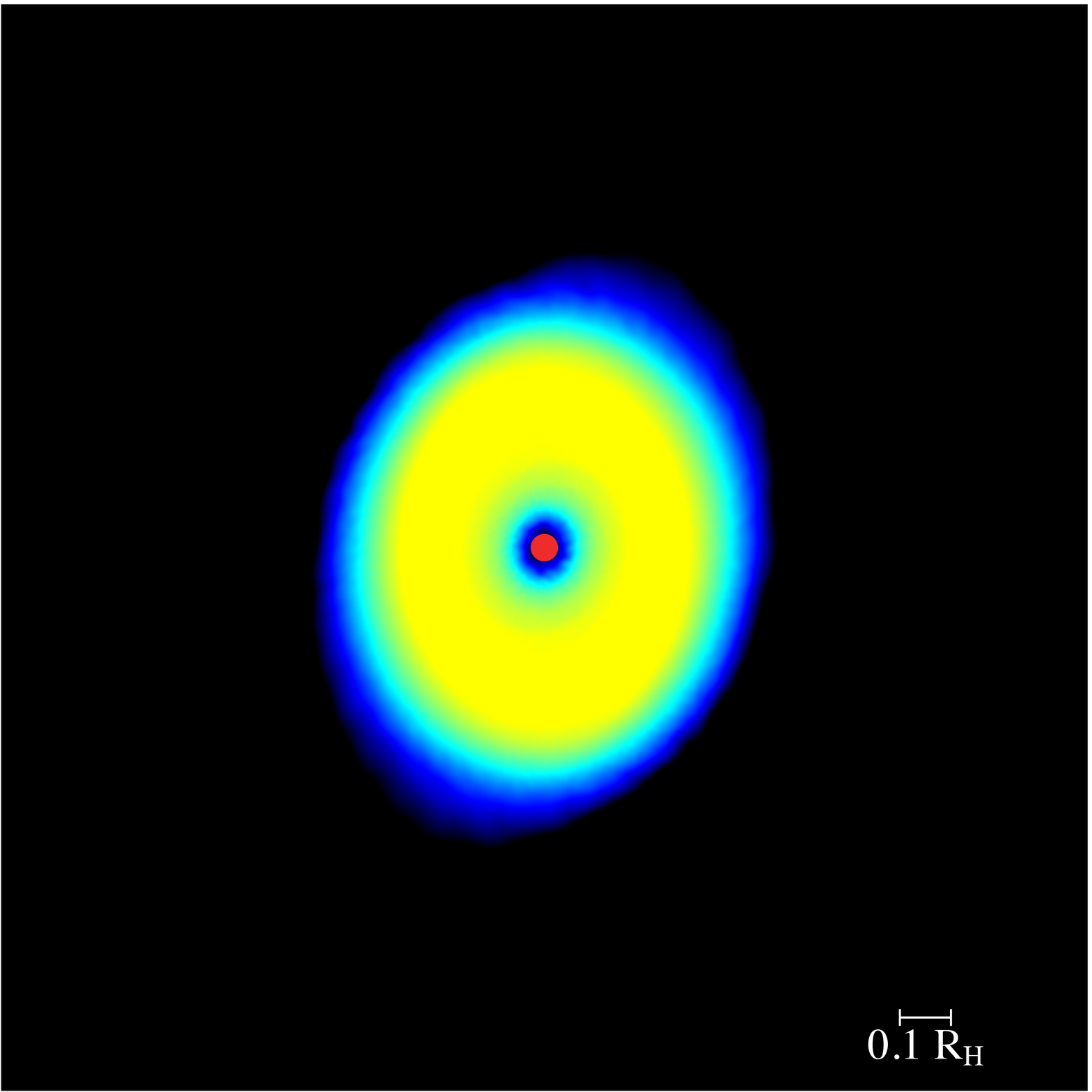} 
\includegraphics[width=5.6cm]{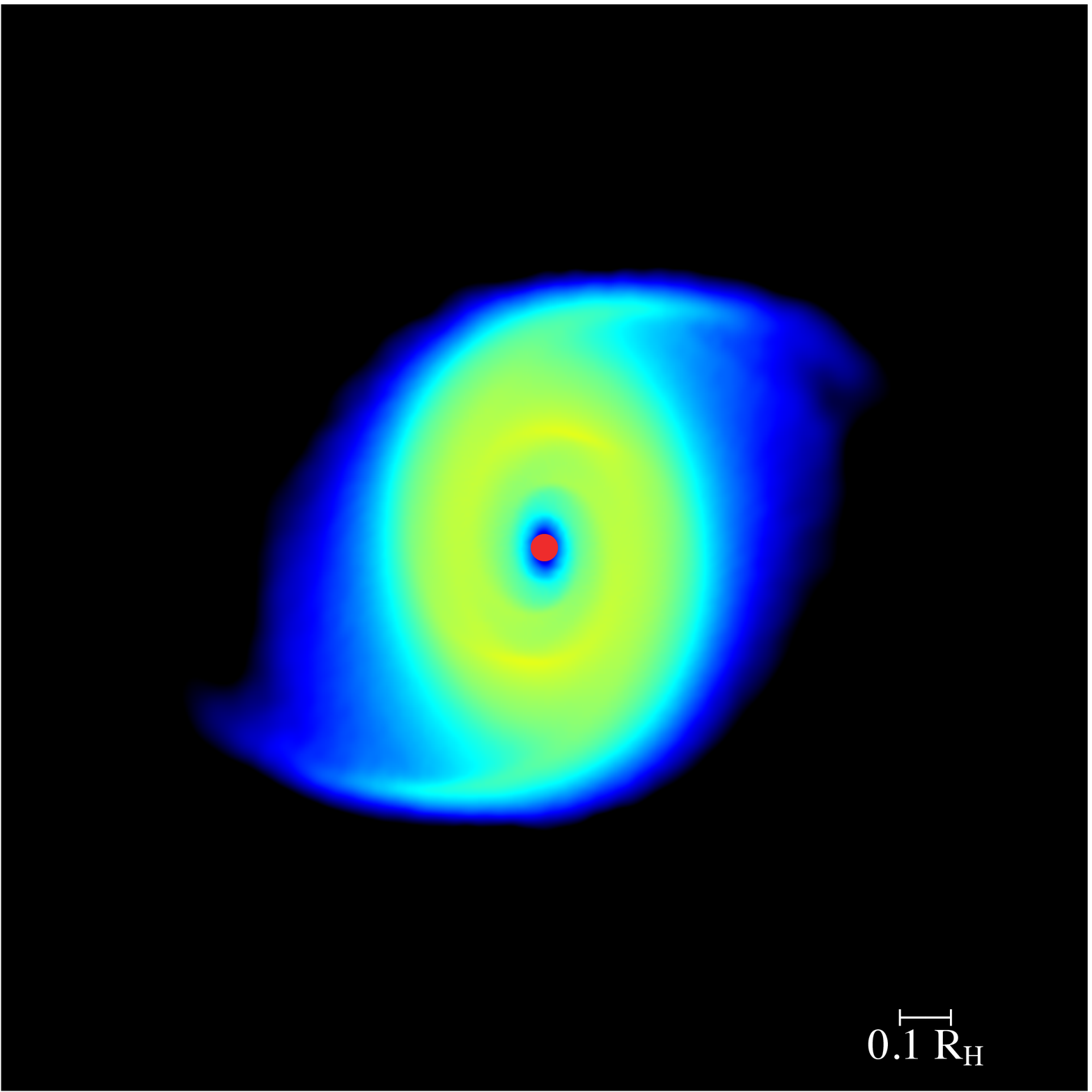} 
\end{centering} 
\caption{The upper panels are the same as Fig.~\ref{fig:inc10} except
  the initial inclination is $i_0=30^\circ$ with disc aspect ratios of
  $(H/r)_{\rm out}=0.025$ (left, run5), $(H/r)_{\rm out}=0.05$
  (middle, run6) and $(H/r)_{\rm out}=0.1$ (right, run7). The lower
  panels show each disc at a time of $18\,P_{\rm orb}$. The view is
  the $x-y$ plane in which the planet orbits. In each panel, the red
  circle shows the planet with its size scaled to the size of the
  accretion radius. The star is along the positive $x$ axis but is not
  shown on this scale. The colour of the gas denotes the column
  density with yellow being about two orders of magnitude larger than
  blue.  }
\label{fig:lowinc} 
\end{figure*}

Fig.~\ref{fig:lowinc} shows simulations with a higher initial
inclination of $i_0=30^\circ$. Similarly to the mildly misaligned
discs, the tilt of the two larger disc aspect ratio discs increase
while the tilt of the disc with lowest aspect ratio decreases. The
tilt of the disc with the largest aspect ratio increases rapidly until
it becomes unstable to KL oscillations above an inclination of about
$60^\circ$. During the KL oscillation the disc eccentricity and
inclination are exchanged. The lower panels of Fig.~\ref{fig:lowinc} show the disc in each simulation at a time of $18\,P_{\rm orb}$. In each case the disc is tilted but there is no eccentricity growth at this time.

The tilt growth timescales for these simulations are shown by
the dotted lines in Fig.~\ref{doublingtimeplot}. The three different
disc aspect ratios are $(H/r)_{\rm out}=0.025$ (run5, black),
$(H/r)_{\rm out}=0.05$ (run6, red) and $(H/r)_{\rm out}=0.1$ (run7,
blue).  The tilt growth timescale is sensitive to the disc
aspect ratio and weakly dependent on the initial disc tilt. The
  tilt growth timescale increases with initial disc tilt.

The accretion rates for these simulations are shown in the dotted
lines in Fig.~\ref{acc}. The accretion rates are very similar to those
of the lower initial inclination discs (solid lines). However, when
the disc undergoes a KL oscillation, the disc eccentricity leads to a
higher accretion rate and also a higher rate of ejection of particles
from the disc \citep[e.g.][]{Franchini2019}. We cannot run the
simulations for longer than shown as we lose resolution in the
disc. In order to explore the behaviour further we next consider an even
higher initial disc misalignment.

\begin{figure*} 
\begin{centering} 
\includegraphics[width=5.6cm]{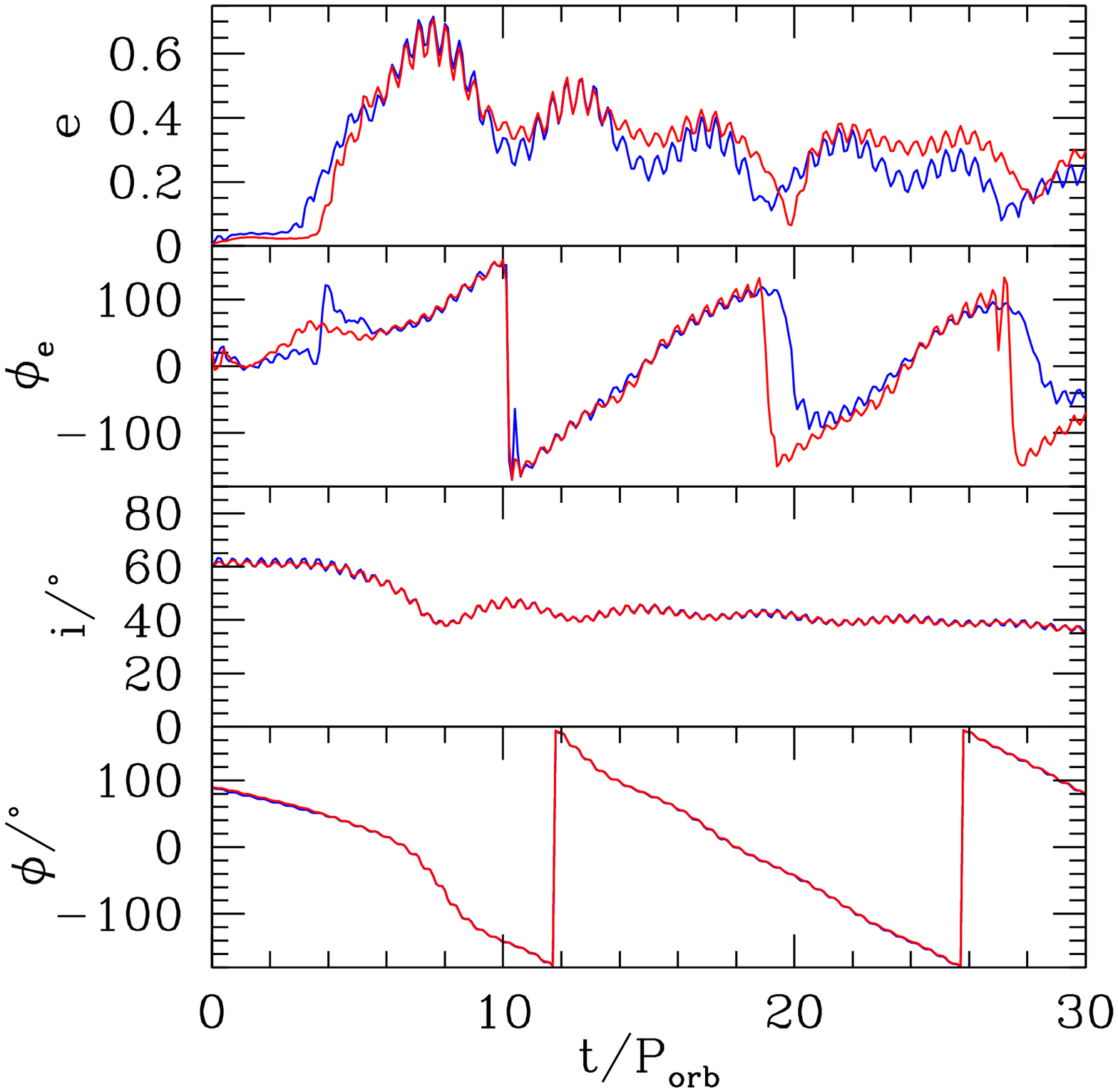} 
\includegraphics[width=5.6cm]{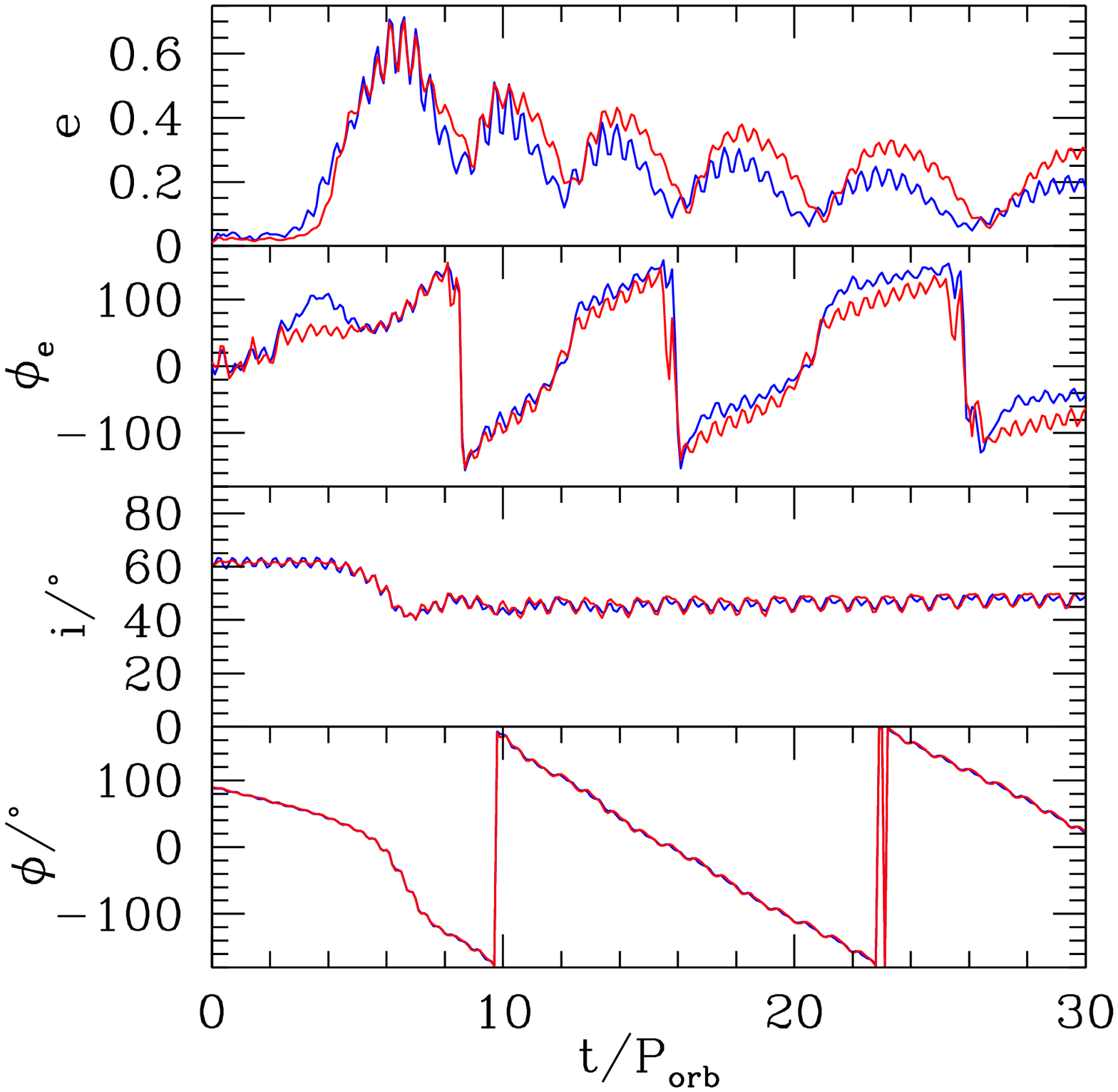} 
\includegraphics[width=5.6cm]{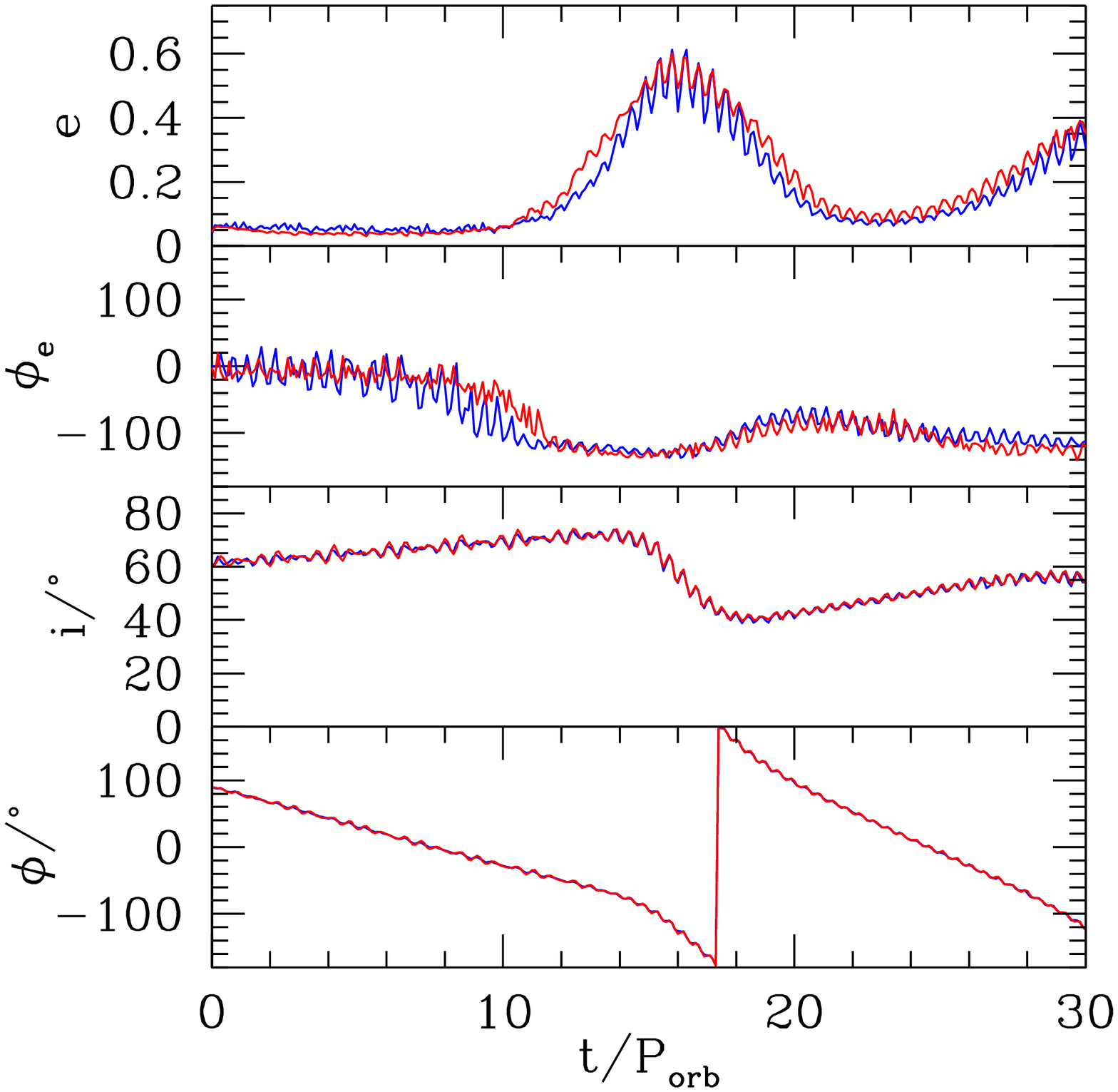} 
\includegraphics[width=5.6cm]{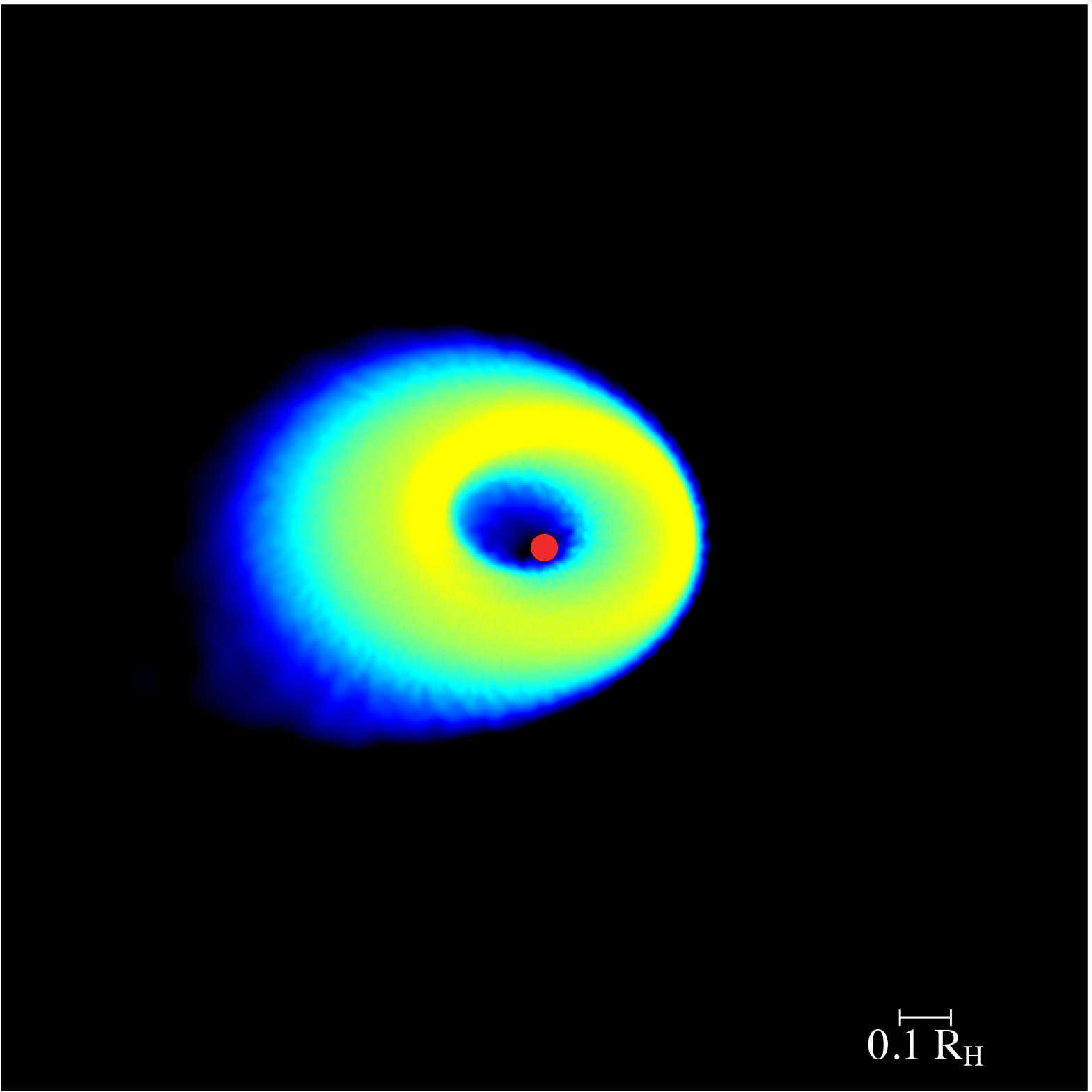} 
\includegraphics[width=5.6cm]{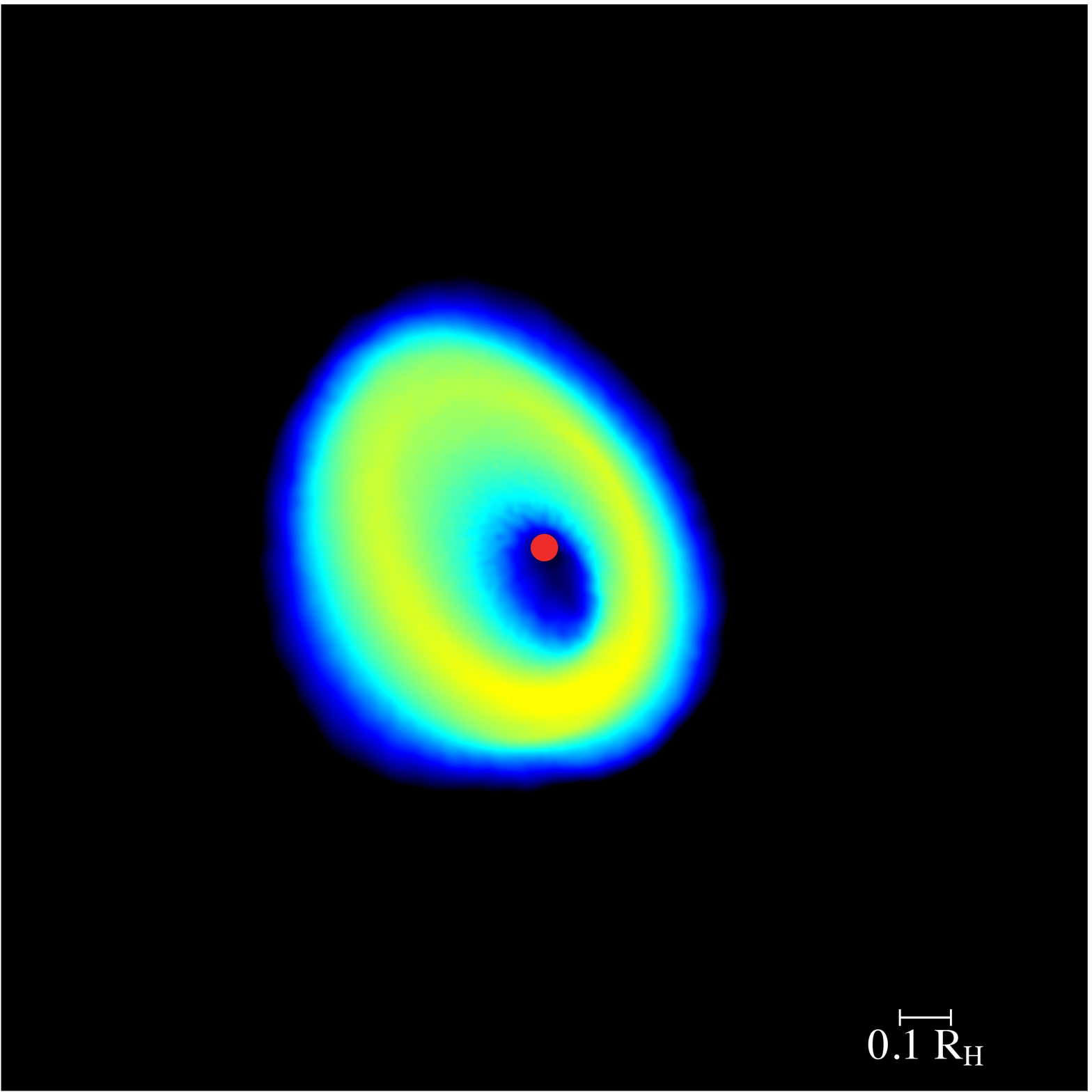} 
\includegraphics[width=5.6cm]{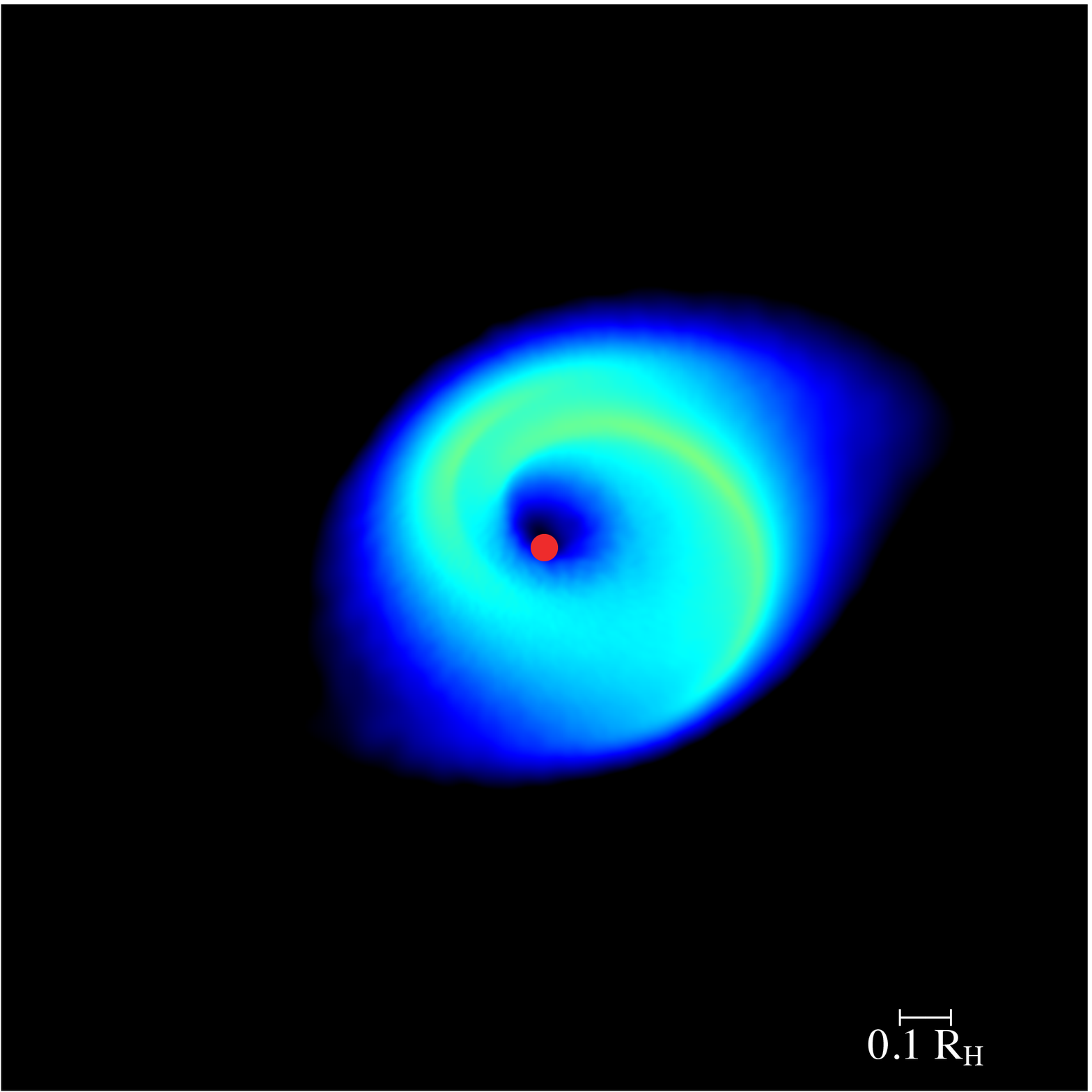} 
\end{centering} 
\caption{Same as Fig.~\ref{fig:lowinc} except the initial disc
  inclination is $i_0=60^\circ$ with disc aspect ratio of $(H/r)_{\rm
    out}=0.025$ (left, run8), $H/r=0.05$ (middle, run9) and
  $H/r=0.1$ (right, run10). }
\label{fig:highinc} 
\end{figure*}

\subsection{Highly inclined circumplanetary discs}
\label{high}

Fig.~\ref{fig:highinc} shows the evolution of discs that begin with an even
higher initial inclination of $i_0=60^\circ$. The discs with lower
disc aspect ratio are already unstable to KL oscillations at this
inclination. The KL oscillations lead to a relatively quick alignment
of the disc to an inclination of about $40^\circ$. This is a result of
dissipation within the disc due to the viscosity and because shocks form during the
oscillations \citep{Martinetal2014b}.  A disc undergoes KL oscillations above a critical inclination that depends upon the disc aspect
ratio \citep{Lubow2017, Zanazzi2017}. The largest disc aspect ratio,
$(H/r)_{\rm out}=0.1$, is not initially unstable at this
inclination. However, since the inclination increases due to the tilt
instability, it becomes unstable later in the simulation. During the
KL oscillation the inclination of the disc decreases. The disc becomes
stable again. The tilt instability again operates increasing the tilt
until it becomes unstable again as seen by the eccentricity growth
near the end of the simulation. Thus, the tilt instability is able to
  increase the range of initial disc tilts that are unstable to KL
  oscillations.

The tilt growth timescale of the simulation with $(H/r)_{\rm out}=0.1$
(run10) is shown in the blue dot-dashed line in
Fig.~\ref{doublingtimeplot}, until the time that the disc becomes KL
unstable. The growth timescale is similar to the discs at lower
inclination with the same disc structure (the other blue lines).   The growth timescale for the simulations with lower $(H/r)_{\rm out}$ (run8 and run9) are not shown because they already undergo KL oscillations from the start.

The
accretion rates on to the planets in the highly misaligned simulations
are shown in the dot-dashed lines in Fig.~\ref{acc}. The KL
oscillations lead to a significantly higher accretion rate on to the
planet during the highly eccentric disc phase.

The simulations show that the nodal phase angles remain roughly
constant with disc radius during the precession.  However, the argument
of periapsis varies significantly over the radial extent of the
disc. This is also clear in the lower panels of
Fig.~\ref{fig:highinc}. The eccentricity vector of the inner hole in
each case is not in the same direction as the eccentricity vector of
the outer parts of the disc. This behaviour was seen before in
protoplanetary disc simulations
\citep[e.g.][]{Martinetal2014b,Fu2015,Fu2015b,Franchini2019}.





\subsubsection{Effect of disc viscosity}

Modelling of protoplanetary discs has been used to infer values of $\alpha$ in the range $10^{-4}-10^{-2}$ \citep[e.g.][]{Hartmann1998,Andrews2009,Hueso2005,Flaherty2015,Rafikov2016,Pinte2016,Flaherty2017,Ansdell2018,Flaherty2018,Flaherty2020}. For circumplanetary discs similar observations are lacking, and the only current constraints are derived from models of satellite formation \citep[e.g.][]{Canup2002,Lubow2013}. However, in general terms we expect that both circumplanetary discs and protoplanetary
discs may contain a dead zone, a region where the magnetorotational
instability (MRI) is inefficient at generating turbulence \citep{Gammie1996}. The corresponding $\alpha$  may therefore be 
low in circumplanetary discs \citep{Lubow2012,Lubow2013,Fujii2014}. The total rate of angular momentum transport in circumplanetary discs may, however, be larger than that in protoplanetary discs because spiral shocks transport angular momentum with an effective $\alpha$ parameter in the range 0.001-0.02 \citep{Zhu2016}. 

\begin{figure} 
\centering
\includegraphics[width=5.6cm]{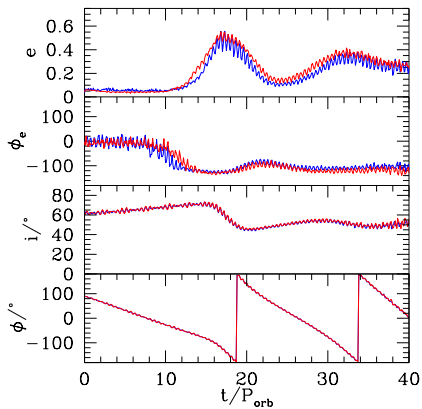} 
\caption{Same as the right panel of Fig.~\ref{fig:highinc} (run10)
  except $\alpha=0.005$ (run11).  }
\label{fig:alpha} 
\end{figure}

All of the simulations considered so far have assumed that
$\alpha=0.01$. To test the sensitivity of the results to this assumption we also consider a simulation with a lower viscosity parameter, $\alpha=0.005$ 
(run11). The lower viscosity leads to a longer viscous
timescale and so we can follow the disc evolution for longer with
better resolution. Fig.~\ref{fig:alpha} shows the evolution of the
disc. The dynamics of the disc are not strongly affected by $\alpha$
as seen by comparing this to the right hand panel of
Fig.~\ref{fig:highinc} (run10) that has $\alpha=0.01$ and otherwise
the same parameters. The green dot-dashed lines in Fig.~\ref{acc} show
the accretion rate on to the planet and the disc mass for this
simulation. The accretion rate is initially slightly lower than that of
the higher $\alpha$ simulation. During the high eccentricity phase of
the KL oscillation, the accretion rate peak is similar. Thus the
effect of lowering $\alpha$ is to increase the lifetime of the disc. The disc is approaching a quasi-steady tilt that is close to the critical angle required for KL disc oscillations.  

\section{Long term evolution of the system}
\label{longterm}

In this Section we discuss some effects that we have neglected in our simulations. We consider the effect of accretion on to the circumplanetary disc from the circumstellar disc, self-gravity of the circumplanetary disc  and the  spin-axis evolution of the planet.

\subsection{Accretion on to the circumplanetary disc}

We have neglected the effect of accretion on to the circumplanetary disc from the circumstellar disc in all the simulations presented in this work. We  estimate the effect of an accretion rate on to the disc, $\dot M$, by calculating the accretion timescale 
\begin{equation}
    t_{\rm acc}=\frac{M_{\rm d}}{\dot M}
\end{equation}
\citep[e.g.][]{Pringle1981}.
For typical parameters we find
\begin{equation}
    \frac{t_{\rm acc}}{P_{\rm orb}}=84 
    \left(\frac{M_{\rm d}}{10^{-6}\,\rm M_\odot}\right)
    \left(\frac{M_{\rm s}}{1\,\rm M_\odot}\right)^{1/2}
     \left(\frac{a}{5.2\,\rm au}\right)^{-3/2}
    \left(\frac{\dot M}{10^{-9}\,\rm M_\odot \, yr^{-1}}\right)^{-1}.
\end{equation}
For these parameters, the accretion timescale is longer than the tilt timescale that is typically a few 10s of planet orbital periods. Once the disc is tilted, the accretion of material may occur at lower tilt angles and thus the tilt of the disc may be damped. We expect the disc to remain unstable to tilting but the timescale for the tilting may be longer than shown in our simulations. The accretion of circular orbit material into an eccentric disc that is undergoing KL oscillations does not necessarily suppress the instability \citep[e.g.][]{Smallwood2020}.

\subsection{Circumplanetary disc self gravity}

 If a circumplanetary disc had sufficient mass to be self-gravitating, the dynamics of the disc may be changed. Strong self-gravity may be able to suppress KL disc oscillations \citep{Batygin2011,Batygin2012,Fu2017}. The \cite{Toomre1964} parameter is
\begin{equation}
    Q=\frac{c_{\rm s}\kappa}{\pi G \Sigma},
\end{equation}
where $c_{\rm s}=H\Omega$, $\Omega=\sqrt{GM_{\rm p}/r^3}$ and $\kappa \approx \Omega $ is the epicyclic frequency. We take the initial surface density profile in our simulations $\Sigma \propto r^{-3/2}$ with  outer disc radius $r_{\rm out}=0.4\, r_{\rm H}$. The Toomre parameter is smallest at the outer disc edge. For $(H/R)_{\rm out}=0.1$, we find that for $Q<1.5$ the circumplanetary disc mass must be greater than $0.2\,M_{\rm p}$. Such a large disc mass may be prohibited from building up because spiral shocks transport angular momentum and enhance accretion on to the planet \citep{Zhu2016,Chen2021}. Thus, it is unlikely that KL oscillations in a circumplanetary disc can be suppressed by self-gravity. 

\cite{Fu2017} found that a  disc that undergoes KL oscillations may be able to fragment even though the disc is Toomre stable in the absence of KL oscillations.  KL circumplanetary disc oscillations may be able to trigger the formation of satellitesimals. Gravitational instability within a protoplanetary disc is an alternative theory for giant planet formation in protoplanetary disc \citep{Boss1997} and may have similar implications for satellite formation in a circumplanetary disc. Gravitational instability can lead to disc fragmentation although the exact conditions required are still being actively researched \citep[e.g.][]{Papaloizou1991,Gammie2001,Lodato2011,Paardekooper2011,Baehr2015,Kratter2016}.   

\subsection{Planet spin axis evolution}

The spin axis of a planet may altered through two effects. First, it may change directly through misaligned accretion on to the planet from the inner edge of the circumplanetary disc. The angular momentum of the accreted material adds to the angular momentum of the planet. Since the mass of the circumplanetary disc is small compared to the mass of the planet, we expect this effect to be small. Secondly, the torque that a tilted disc exerts on a spinning oblate planet may change the planet spin-axis. This has been discussed before in the black hole case where the torque of a tilted disc aligns the black hole and the disc on a timescale that is much faster than accretion alone \citep[e.g.][]{SF1996,Martinetal2007,Martinetal2009}. We will investigate this in more detail in a  future publication.

\section{Discussion}
\label{discussion}

The interplay of tilt and KL instability of circumplanetary discs has several
potential implications. 
For the tilt mechanism to operate we have shown that the disc aspect
ratio must be relatively large. This is expected for
circumplanetary discs provided that the mass of the planet is lower than a threshold value. For a disc in a steady state, the disc aspect
ratio scales with the planet mass as $H/r\propto M_{\rm p}^{-1/3}$, with a 
weaker positive dependence on the planetary accretion rate 
\citep[see equation 4 in][]{Martin2011}. Thus, the larger the planet
mass, the smaller the disc aspect ratio. In the Solar System, different dynamics could therefore occur for Jupiter when compared to Saturn, Uranus and Neptune. We note that Jupiter has an
equatorial plane that is close to aligned to its orbit, while the
remaining giant planets are more
highly misaligned.

Since regular satellites are thought to form in a circumplanetary
disc, the eccentricity of the disc would affect their formation. In the protoplanetary disc case eccentricity due to the interaction
of a close and eccentric binary companion may lead to destructive planetesimal collisions \citep{Paardekooper2008,Kley2008,Paardekooper2010,Marzari2012}, hindering the the formation of
planets \citep{Rafikov2013,Rafikov2015,Rafikov2015b,Silsbee2015}. 
The presence of the gas disc can counteract this dynamical effect by providing a drag that is able to align
the orbital eccentricities of the planetesimals, reducing the
relative speeds of collisions \citep{Marzari2000}. The
eccentricities experienced in a circumplanetary disc undergoing KL
oscillation may be high, $e \gtrsim 0.5$. Regular satellite formation
may need to proceed through different mechanisms in such highly eccentric discs. We also note that the threshold conditions for KL oscillations differ for gas discs and purely N-body systems. 
Solid bodies that form within a circumplanetary gas disc that is stable against KL oscillations  may become unstable once the gas disc has dissipated if their inclination
is above $39^\circ$ \citep[e.g.][]{Speedie2020}.

It is as yet unclear what will prove to be the best observational diagnostics of circumplanetary discs. Kinematic tracers provide powerful tools for characterizing eccentricities and misalignments in protoplanetary discs \citep[e.g.][]{Bi2020}, and for testing the hypothesis that spirals, rings, and cavities within protoplanetary discs are caused by planets \citep{Muzerolle2010,Zhu2011,Zhang18}. With sufficient sensitivity and spatial resolution, the distinctive kinematics of circumplanetary discs experiencing a phase of KL oscillations may be detectable. Currently, the indirect evidence pointing to a large disc-embedded planet population at large orbital radii is in some tension with direct imaging surveys that tend to suggest a 
smaller population, at least at high masses \citep[e.g.][]{Bowler2016}. 
Episodic
accretion through a circumplanetary disc may be the cause of the  low
detection rate of accreting planets  \citep{Brittain2020}. A
circumplanetary disc that is undergoing KL oscillations would indeed
undergo episodic accretion as shown in Fig.~\ref{acc}.

\section{Conclusions}
\label{concs}

Fluid discs around one component of a binary system that are sufficiently misaligned can be unstable to global KL disc
oscillations 
\citep{Martinetal2014b}. Furthermore, circumplanetary discs are
linearly unstable to the growth of a disc tilt \citep{Lubow92,
  Lubow2000, MZA2020}. In this paper we have shown that detached
circumplanetary discs have favourable disc aspect ratios to become KL
unstable after an initial phase of inclination growth due to the tilt
instability.  We have neglected accretion on to the
circumplanetary disc that we expect to damp the tilt growth. However,
for high enough disc aspect ratio, $H/r\gtrsim 0.05$,  circumplanetary discs  become
misaligned on a short timescale and, provided that $\alpha$ is
sufficiently small and the disc lasts long enough, become KL
unstable. 
A planet with a circumplanetary disc that is undergoing KL oscillations has episodic accretion that may explain the low detection rate of accreting planets.  
The eccentricity of the disc also has implications for regular
satellite formation in the disc and the observability of
circumplanetary discs and forming planets.

\section*{Data availability statement}
The results in this paper can be reproduced using the {\sc phantom} code (Astrophysics Source Code Library identifier {\tt ascl.net/1709.002}). The data underlying this article will be shared on reasonable request to the corresponding author.

\section*{Acknowledgements} 
We thank Daniel Price for providing the {\sc phantom} code for SPH
simulations and acknowledge the use of SPLASH \citep{Price2007} for
the rendering of the figures. Computer support was provided by
UNLV's National Supercomputing Center. We acknowledge support from
NASA TCAN award 80NSSC19K0639. CCY is grateful for the support from NASA via the Emerging Worlds program (Grant 80NSSC20K0347).
 
\bibliographystyle{mnras}

\label{lastpage} 
\end{document}